# Spatiotemporal Flexible Sparse Reconstruction for Rapid Dynamic Contrast-enhanced MRI

Yuhan Hu, Xinlin Zhang, Li Feng, Dicheng Chen, Zhiping Yan, Xiaoyong Shen, Gen Yan, Lin Ou-yang, Xiaobo Qu*

*Abstract*—Dynamic Contrast-enhanced magnetic resonance imaging (DCE-MRI) is a tissue perfusion imaging technique. Some versatile free-breathing DCE-MRI techniques combining compressed sensing (CS) and parallel imaging with golden-angle radial sampling have been developed to improve motion robustness with high spatial and temporal resolution. These methods have demonstrated good diagnostic performance in clinical setting, but the reconstruction quality will degrade at high acceleration rates and overall reconstruction time remains long. In this paper, we proposed a new parallel CS reconstruction model for DCE-MRI that enforces flexible weighted sparse constraint along both spatial and temporal dimensions. Weights were introduced to flexibly adjust the importance of time and space sparsity, and we derived a fast thresholding algorithm which was proven to be simple and efficient for solving the proposed reconstruction model. Results on *in vivo* liver DCE datasets show that the proposed method outperforms the state-of-the-art methods in terms of visual image quality assessment and reconstruction speed without introducing significant temporal blurring.

## I. Introduction

MAGNETIC resonance imaging (MRI) is a non-invasive, radiation-free modality that plays an essential role in day-to-day routine clinical diagnosis.

This work was supported in part by National Key R&D Program of China (2017YFC0108703), National Natural Science Foundation of China (61971361, 61871341, 61671399, and 61811530021), Natural Science Foundation of Fujian Province of China (2018J06018), Fundamental Research Funds for the Central Universities (20720180056), Science and Technology Program of Xiamen (3502Z20183053), Xiamen University Nanqiang Outstanding Talents Program. (*Corresponding author: Xiaobo Qu with e-mail: quxiaobo@xmu.edu.cn).

Yuhan Hu, Xinlin Zhang, Dicheng Chen, and Xiaobo Qu are with Department of Electronic Science, Fujian Provincial Key Laboratory of Plasma and Magnetic Resonance, School of Electronic Science and Engineering, National Model Microelectronics College, Xiamen University, Xiamen 361105, China.
Li Feng is with Department of Medical Physics Memorial Sloan Kettering Cancer Center, 1250 1st Ave, New York, NY 10065.
Zhiping Yan is with Department of Fujian Medical University Xiamen Humanity Hospital.
Xiaoyong Shen is with Department of The First Affiliated Hospital of Zhejiang University School of Medicine.
Gen Yan is with Department of Radiology, The Second Hospital of Xiamen Medical College.
Lin Ou-yang is with Department of Medical Imaging of Southeast Hospital, Medical College of Xiamen University, Zhangzhou 363000, China.

In recent years, remarkable advances have been achieved in dynamic contrast-enhanced magnetic resonance imaging (DCE-MRI) in terms of spatiotemporal resolution, image quality and motion management. DCE-MRI is a tissue perfusion technique and allows quantitative characterization of the microcirculation and tissue characteristics through analysis of signal intensity changes following injection of a contrast agent. It has been an integral part in most routine clinical MRI protocols for detection of suspected lesions and for evaluation treatment response [1-5].

Fast data acquisition speed is needed to capture changes in signal intensity as the contrast-agent passes through the cardiovascular system [6] and to ensure adequate spatial and temporal resolution. To accelerate DCE-MRI, a variety of methods have been proposed, including various parallel imaging methods (e.g. SENSE [7] and GROWL [8]), golden-angle radial sampling [9], k-t acceleration methods (e.g. radial K-T SPIRiT [10]), artificial sparsity methods (e.g. ARTS-GROWL [11] and K-T ARTS-GROWL [12]), the combination of parallel imaging, golden-angle radial sampling and compressed sensing method, iGRASP [13].

The principle of compressed sensing (CS) [14-20] meets the need of accelerating DCE-MRI. The key point for developing CS-based reconstruction methods is the sparse representation of images. Sparse representation can be divided into two main categories: orthogonal sparse representation system [16, 21, 22] and redundant sparse representation system [23-27]. While the former system is helpful for theoretical analysis, fast algorithm design, reducing calculation time and memory consumption, it often leads to insufficient sparse representation of the image. The latter, on the other hand, is able to capture more image features, thereby better eliminating noise and suppressing residual artifacts [28, 29]. There are two different models for MRI image reconstruction called synthetic model and analysis model under redundant sparse representation system which is represented mainly by tight frames [27], and it has been shown that the analysis model can achieves better image quality [28, 30, 31]. To solve the analysis model, our group proposed a projected iterative soft-thresholding algorithm (pISTA) and its acceleration version - pFISTA [28] by rewriting the analysis model into an equivalent synthesis-like one and calculating the proximal map of non-smooth sparsity terms approximately. Comparing with other state-of-the-art algorithms that can solve the analysis model, such as ADMM [32] and SFISTA [33], pFISTA consumes low memory and only needs to tune one free parameter with simple settings [28].

In DCE-MRI reconstruction, we need to preserve temporal fidelity, which can be assessed by dynamic enhancement curve (signal intensity changes over time) [12, 13] firstly, and then seek to improve the image quality as much as possible (less artifacts and/or noise). This poses a major challenge for developing reconstruction model based on CS.

The existing CS-based reconstruction methods in DCE-MRI, such as iGRASP, only constrain sparsity in the time dimension. In this work, we proposed a parallel dynamic analysis reconstruction model with sparsity constraints in both the time and spatial dimension. Weights were introduced to flexibly adjust the balance of the sparsity between two dimensions, and pFISTA was modified to solve the model. We have also shown the convergence results of applying pFISTA for reconstructing DCE-MRI images.

The rest of the paper is organized as follows. In Section II, we introduced prior related works including continuous 3D data acquisition scheme, iGRASP, K-T ARTS-GROWL, and pFISTA. In Section III, we introduced the proposed model first then derive the numerical algorithm, and analyze the convergence. In Section IV, the performance of the proposed method is demonstrated by experiments on various *in vivo* liver DCE datasets. Finally, we concluded this paper in section V.

## II. RELATED WORK

### A. Free-breathing continuous 3D data acquisition scheme

To implement continuous 3D data acquisition, cartesian sampling along the partition dimension ($k_z$) and golden-angle radial sampling (radial lines are continuously increased by 111.25 degrees) in $k_x$-$k_y$ plane are combined together (Fig.1) [13]. Radial sampling has a lower sensitivity to motion. Moreover, approximately uniform k-space coverage can be obtained for any arbitrary number of consecutive spokes. Therefore, in this work, certain Fibonacci numbers which can obtain an optimal SNR [34] of consecutive spokes are combined to form each time frame.

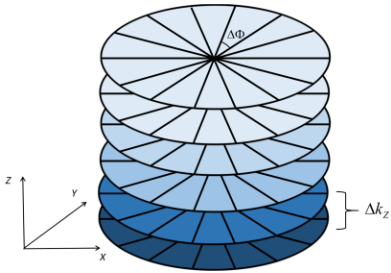

Fig. 1. Free-breathing continuous 3D data acquisition scheme.

### B. Iterative Golden-angle Radial Sparse Parallel MRI

iGRASP [13] combines parallel imaging, golden-angle radial sampling and CS. To reconstruct DCE image series, iGRASP firstly sorts the golden-angle radial k-space data into dynamic series by grouping a number of consecutive Fibonacci spokes into each temporal frame. Iterative reconstruction is then performed on the undersampled re-sorted radial data using the first order finite differences as a temporal sparsity transform. The reconstruction process can be expressed as Eq. (1). The original iGRASP implementation uses the nonlinear conjugate gradient algorithm (CG) [16] to solve the reconstruction problem.

$$\hat{\mathbf{d}} = \arg\min \left\{ \left\| \mathbf{F} \cdot \mathbf{S} \cdot \mathbf{d} - \mathbf{m} \right\|_2^2 + \lambda \left\| \mathbf{T} \cdot \mathbf{d} \right\|_1 \right\}, \quad (1)$$

where $\hat{\mathbf{d}}$ is the image series to be reconstructed in x-y-t space, $\mathbf{T}$ is first order finite difference operator along the time dimension, $\mathbf{m}$ is the undersampled k-space data, $\mathbf{S}$ is the coil sensitivity maps, $\mathbf{F}$ is the nonuniform fast Fourier transform operator (NUFFT) [35], and $\lambda$ is the regularization parameter.

### C. K-T ARTS-GROWL

K-T ARTS-GROWL [12] is an extension of ARTS-GROWL [11]. This technique is comprised of three steps. The first step aims to use GROWL [8] to obtain an intermediate parallel imaging results, followed by the second step apply a K-T sparse denoising method to denoise the parallel imaging results. The reconstruction problem can be expressed as Eq. (2), which is solved using the CG algorithm [16].

$$\hat{\mathbf{I}} = \arg\min_{\mathbf{I}} \left\| \mathbf{I} - \mathbf{I}_p \right\|_2^2 + \lambda \left\| \mathbf{TI} \right\|_1, \quad (2)$$

where $\hat{\mathbf{I}}$ is the denoised image series in x-y-t space, $\mathbf{I}_p$ is the parallel imaging results, $\mathbf{I}$ is the output image series, $\mathbf{T}$ is first order finite difference operator along the time dimension, and $\lambda$ is the regularization parameter.

Finally, the sparse reconstruction results and K-T denoised image series were summed up to obtain the final result.

### D. Projected fast iterative soft-thresholding algorithm

An analysis sparse reconstruction model, in which a MRI image is under a sparse representation of tight frame, can be expressed as:

$$\min_{\mathbf{x}} \frac{1}{2} \left\| \mathbf{y} - \mathbf{UFx} \right\|_2^2 + \lambda \left\| \mathbf{\Psi x} \right\|_1, \quad (3)$$

where $\mathbf{x}$ is the MRI image data rearranged to a column vector, $\mathbf{y}$ is the undersampled k-space data, $\mathbf{F}$ is the discrete Fourier transform and $\mathbf{U}$ is the undersampling matrix, $\mathbf{\Psi}$ is a tight frame to sparsify an image, and $\lambda$ is the regularization parameter.

To solve the model in Eq. (3), pFISTA rewrites the analysis model above into a synthesis-like model as:

$$\min_{\mathbf{\alpha} \in Range(\mathbf{\Psi})} \frac{1}{2} \left\| \mathbf{y} - \mathbf{UF\Psi}^* \mathbf{\alpha} \right\|_2^2 + \lambda \left\| \mathbf{\alpha} \right\|_1, \quad (4)$$

where $\mathbf{\Psi}^*$ is the adjoint of the tight frame $\mathbf{\Psi}$, and it specifically satisfies $\mathbf{\Psi}^* \mathbf{\Psi} = \mathbf{I}$. $\mathbf{\alpha}$ contains the sparse coefficients of an image under the representation $\mathbf{\Psi}^*$.

The iterations of pFISTA to solve the synthesis-like model in Eq. (4) is [28]:

$$\mathbf{x}_{k+1} = \boldsymbol{\Psi}^* T_{\gamma\lambda}\left(\boldsymbol{\Psi}\left(\hat{\mathbf{x}}_k + \gamma \mathbf{F}^* \mathbf{U}^T\left(\mathbf{y} - \mathbf{UF}\hat{\mathbf{x}}_k\right)\right)\right),$$

$$t_{k+1} = \frac{1+\sqrt{1+4t_k^2}}{2}, \quad (5)$$

$$\hat{\mathbf{x}}_{k+1} = \mathbf{x}_{k+1} + \frac{t_k - 1}{t_{k+1}}(\mathbf{x}_{k+1} - \mathbf{x}_k),$$

where $\gamma$ is the step size, $T_{\gamma\lambda}(\cdot)$ is a point wise soft-thresholding function defined as:

$$T_{\gamma\lambda}(\alpha_i) = \max\{|\alpha_i| - \gamma\lambda, 0\} \cdot \frac{\alpha_i}{|\alpha_i|}. \quad (6)$$

## III. PROPOSED METHOD

### A. Reconstruction model and pipeline

To achieve a desired temporal resolution, the continuously acquired golden-angle radial spokes are sorted into temporal frames by grouping a Fibonacci number of spokes to form each frame. To reconstruct the image series, tight frames in time and spatial dimensions are applied separately to sparsify the image series, which leads to the $l_1$ norm optimization problem. To accomplish the goal of that, the importance of time and spatial sparsity can be flexibly adjusted and corresponding weights are introduced into the $l_1$ norm-based problem. The proposed model can be expressed as Eq. (7) and corresponding reconstruction pipeline is shown in Fig. 2.

$$\min_{\mathbf{d}} \lambda \left\| \begin{bmatrix} w_t \mathbf{R}_T \mathbf{d} \\ w_s \mathbf{R}_S \mathbf{d} \end{bmatrix} \right\|_1 + \frac{1}{2}\|\mathbf{Ed} - \mathbf{m}\|_2^2, \quad (7)$$

where $\mathbf{d}$ is the image series to be reconstructed which is rearranged into a column vector, $\mathbf{m}$ is the undersampled k-space data, $\lambda$ is the regularization parameter to balance the sparsity and data consistency. $\mathbf{E}$ is given by the multiplication of NUFFT encoding elements and coil sensitivities, besides the adjoint operator of $\mathbf{E}$ is $\mathbf{E}^*$ [36]. The required coil sensitivities are estimated from the temporal average of all acquired spokes using the adaptive coil combination technique [13, 37]. $\mathbf{R}_T$ is a tight frame used to sparsify the image series along the time dimension, selected cyclic shift discrete wavelet transform (CSDWT). The reason we chose the CSDWT is to keep temporal fidelity of the reconstruction result while maintaining the tight frame and sparse representation property. $\mathbf{R}_S$ is also a tight frame adopted to sparsify the image series in spatial dimension chosen shift-invariant discrete wavelet transform (SIDWT) [38]. Related details about $\mathbf{R}_T$ (CSDWT), $\mathbf{R}_S$ (SIDWT) can be seen in the Appendix. $w_t = 1$ and $w_s = w$ representing the importance of temporal and spatial sparsity, respectively.

One representative frame (the Venous phase) of the reconstructed image series under different weights is shown in Fig. 3, indicating that the introduced weights are meaningful as we discussed below:

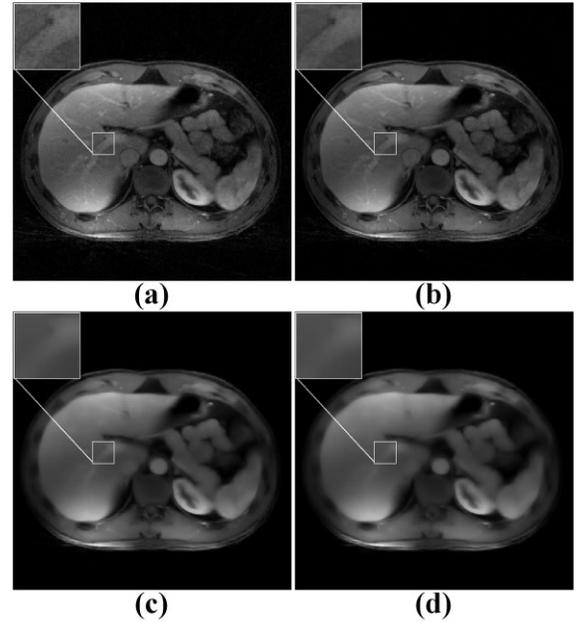

Fig.3. Representative frame of the reconstruction results under different weights. (a): $w_s = 0$, (b): $w_s = 0.09$, (c): $w_s = 1$, (d): $w_s = 1.5$

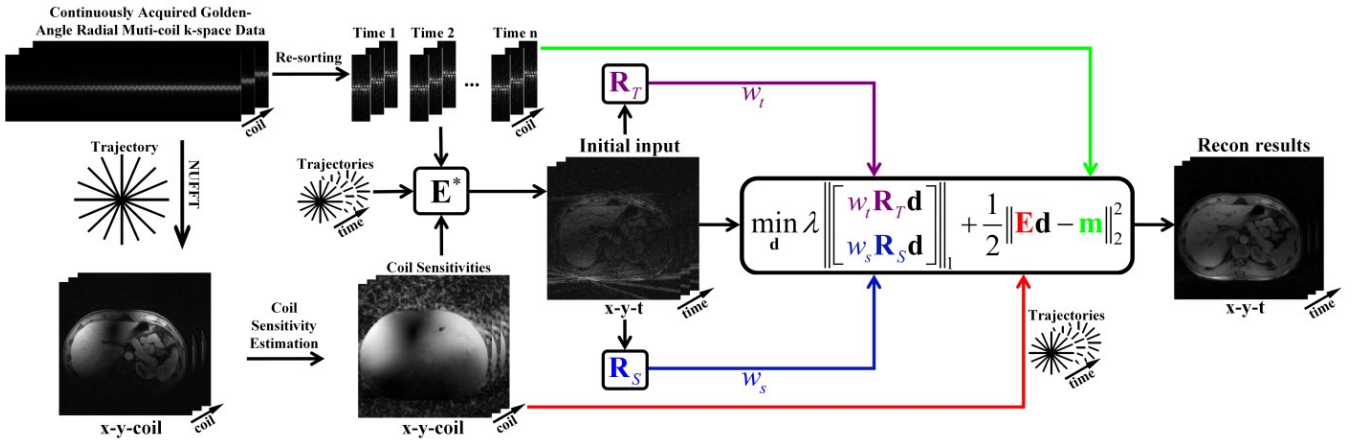

Fig. 2. The proposed method reconstruction pipeline. the continuously acquired data are firstly re-sorted into undersampled dynamic time series by grouping several consecutive spokes. The proposed method is then applied to the re-sorted multi-coil radial data with weighted sparse in time and space dimension constraints, using the operator $\mathbf{E}^*$ to produce the image time-series (x-y-t). Acquired coil sensitivity maps are estimated from the multi-coil reference images (x-y-coil) which are given by the coil-by-coil NUFFT reconstruction of the temporal average of all acquired spokes.

When we enhanced the time sparsity ($w_s = 0.09$, Fig.3(b)), it can get a good result. Without spatial sparsity ($w_s = 0$, Fig.3(a)), the image suffers from an increased level of noise. When we keep the same contribution of time and spatial sparsity ($w_s = 1$, Fig.3(c)), or enhanced the spatial sparsity ($w_s = 1.5$, Fig.3(d)), the image suffers from blurring.

*B. Numerical algorithm and convergence analysis*

For derivation convenience, the proposed model in Eq. (7) is transformed into an equivalent form in Eq. (8).

$$\min_{\mathbf{d}} \lambda \|\mathbf{WAd}\|_1 + \frac{1}{2}\|\mathbf{Ed} - \mathbf{m}\|_2^2, \quad (8)$$

where $\mathbf{A} = \begin{bmatrix} \mathbf{R}_T \\ \mathbf{R}_S \end{bmatrix}$, $\mathbf{W} = diag(w_t, \cdots, w_t, w_s, \cdots, w_s)$ is a diagonal matrix.

The adjoint of $\mathbf{A}$ can be denoted as $\mathbf{A}^*$, which specifically satisfies:

$$\mathbf{A}^*\mathbf{A} = \begin{bmatrix} \mathbf{R}_T^* & \mathbf{R}_S^* \end{bmatrix}\begin{bmatrix} \mathbf{R}_T \\ \mathbf{R}_S \end{bmatrix} = \mathbf{R}_T^*\mathbf{R}_T + \mathbf{R}_S^*\mathbf{R}_S \quad (9)$$

where $\mathbf{R}_T^*$, $\mathbf{R}_S^*$ are the adjoint of tight frames $\mathbf{R}_T$ and $\mathbf{R}_S$, satisfying $\mathbf{R}_T^*\mathbf{R}_T = \mathbf{I}$ and $\mathbf{R}_S^*\mathbf{R}_S = \mathbf{I}$, so we have:

$$\mathbf{A}^*\mathbf{A} = 2\mathbf{I}. \quad (10)$$

The modification helps us to easily introduce the pFISTA to solve the reconstruction model with the following iterations:

$$\mathbf{d}_{k+1} = \frac{1}{2}\mathbf{A}^* T_{\gamma\lambda w_i}\left(\mathbf{A}\left(\hat{\mathbf{d}}_k - \frac{1}{2}\gamma\mathbf{E}^*\left(\mathbf{E}\hat{\mathbf{d}}_k - \mathbf{m}\right)\right)\right),$$

$$t_{k+1} = \frac{1 + \sqrt{1 + 4(t_k)^2}}{2}, \quad (11)$$

$$\hat{\mathbf{d}}_{k+1} = \mathbf{d}_{k+1} + \left(\frac{t_k - 1}{t_{k+1}}\right)(\mathbf{d}_{k+1} - \mathbf{d}_k).$$

where $\gamma$ is the step size, $w_i$ is the weight, $T_{\gamma\lambda w_i}(\cdot)$ is a point wise soft-thresholding function defined as:

$$T_{\gamma\lambda w_i}(\alpha_i) = \max\{|\alpha_i| - \gamma\lambda w_i, 0\} \cdot \frac{\alpha_i}{|\alpha_i|}. \quad (12)$$

Good empirical convergence of the objective function is observed in Fig.4. Next, we theoretically analyze the convergence of pFISTA for DCE-MRI.

**Theorem 1**: let $\{\mathbf{d}_k\}$ be generated by pFISTA, and when the step size satisfies $0 < \gamma \le 1$, the sequence $\{\boldsymbol{\alpha}_k\} = \{\mathbf{Ad}_k\}$ will converge to a solution of:

$$\min_{\boldsymbol{\alpha}} \lambda \|\mathbf{W\alpha}\|_1 + \frac{1}{2}\left\|\frac{1}{2}\mathbf{EA}^*\boldsymbol{\alpha} - \mathbf{m}\right\|_2^2 + \frac{1}{4\gamma}\left\|\left(\mathbf{I} - \frac{1}{2}\mathbf{AA}^*\right)\boldsymbol{\alpha}\right\|_2^2. \quad (13)$$

And the convergence speed is:

$$F(\boldsymbol{\alpha}_k) - F(\boldsymbol{\alpha}^*) \le \frac{2}{\gamma(k+1)^2}\|\boldsymbol{\alpha}_0 - \boldsymbol{\alpha}^*\|, \quad (14)$$

where $\boldsymbol{\alpha}^*$ is a solution of Eq. (13) and $F(\cdot)$ is the objective function in Eq. (13).

Proof of **Theorem 1**:
Denote the gradient term in Eq.(13) as:

$$u(\boldsymbol{\alpha}) = \frac{1}{2}\left\|\frac{1}{2}\mathbf{EA}^*\boldsymbol{\alpha} - \mathbf{m}\right\|_2^2 + \frac{1}{4\gamma}\left\|\left(\mathbf{I} - \frac{1}{2}\mathbf{AA}^*\right)\boldsymbol{\alpha}\right\|_2^2. \quad (15)$$

According to the convergence analysis in [28, 29, 39], the convergence depends on the Lipschitz constant of the gradient $\nabla u$ that is:

$$L(\gamma) = L(\nabla u) = \left\|\frac{1}{4}\mathbf{AE}^*\mathbf{EA}^* + \frac{1}{2\gamma}\left(\mathbf{I} - \frac{1}{2}\mathbf{AA}^*\right)\right\|_2. \quad (16)$$

Based on the results in [28, 29], if the step size satisfies:

$$\gamma \le \frac{1}{L(\gamma)}. \quad (17)$$

or equivalently

$$L(\gamma) \le \frac{1}{\gamma}. \quad (18)$$

The algorithm will converge with a speed described in Eq. (14). Now the question is directly related to analyze $L(\gamma)$.

Let: $\mathbf{B} = \frac{1}{4}\mathbf{AE}^*\mathbf{EA}^* - \frac{1}{4\gamma}\mathbf{AA}^*$, we have:

$$L(\gamma) = \left\|\frac{1}{4}\mathbf{AE}^*\mathbf{EA}^* + \frac{1}{2\gamma}\left(\mathbf{I} - \frac{1}{2}\mathbf{AA}^*\right)\right\|_2 = \left\|\mathbf{B} + \frac{1}{2\gamma}\mathbf{I}\right\|_2. \quad (19)$$

So:

$$L(\gamma) = \left\|\mathbf{B} + \frac{1}{2\gamma}\mathbf{I}\right\|_2 = \max_i\left(\left|c_i(\mathbf{B}) + \frac{1}{2\gamma}\right|\right), \quad (20)$$

where $c_i(B)$ is the $i^{th}$ eigenvalue of $\mathbf{B}$. $\mathbf{B} + \frac{1}{2\gamma}\mathbf{I}$ is a Hermitian matrix, therefore we need to further analyze $c_i(\mathbf{B})$. Noting that $\mathbf{A}^*\mathbf{A} = 2\mathbf{I}$, we can get:

$$\mathbf{Bz} = \left(\frac{1}{4}\mathbf{AE}^*\mathbf{EA}^* - \frac{1}{4\gamma}\mathbf{AA}^*\right)\mathbf{z} = \beta\mathbf{z}, \quad (21)$$

where $\mathbf{z}$ is the eigenvector of $\mathbf{B}$ belonging to the eigenvalue $\beta$.

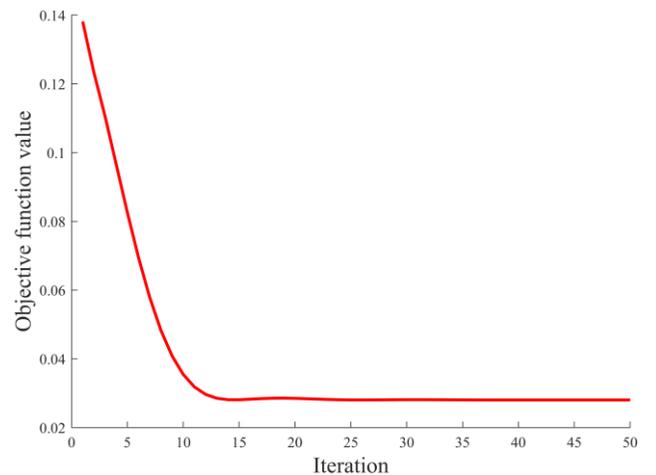

Fig. 4. The changes of objective function value during iterations.

That means all non-zero eigenvalues of $\mathbf{B}$ satisfies:

$$c_i(\mathbf{B}) \in \left\{c_i\left(\frac{1}{2}\mathbf{E}^*\mathbf{E} - \frac{1}{2\gamma}\mathbf{I}\right)\right\} = \left\{c_i\left(\frac{1}{2}\mathbf{E}^*\mathbf{E}\right) - \frac{1}{2\gamma}\right\}. \quad (22)$$

Thus:

$$L(\gamma) = \max_i \left\{\left|c_i(\mathbf{B}) + \frac{1}{2\gamma}\right|\right\} = \max_i \left\{\frac{1}{2\gamma}, \left|c_i\left(\frac{1}{2}\mathbf{E}^*\mathbf{E}\right)\right|\right\}. \quad (23)$$

And with a normalization in which $\|\mathbf{E}\|_2^2 \leq 1$ [36], so we have:

$$\max_i \left|c_i\left(\frac{1}{2}\mathbf{E}^*\mathbf{E}\right)\right| = \frac{1}{2}\|\mathbf{E}^*\mathbf{E}\|_2 \leq \frac{1}{2}\|\mathbf{E}\|_2^2 \leq \frac{1}{2}. \quad (24)$$

Therefore,

$$L(\gamma) = \max_i \left\{\frac{1}{2\gamma}, \left|c_i\left(\frac{1}{2}\mathbf{E}^*\mathbf{E}\right)\right|\right\} = \frac{1}{2\gamma} < \frac{1}{\gamma}, 0 < \gamma \leq 1,$$

$$L(\gamma) = \max_i \left\{\frac{1}{2\gamma}, \left|c_i\left(\frac{1}{2}\mathbf{E}^*\mathbf{E}\right)\right|\right\} \leq \frac{1}{2}, \gamma > 1. \quad (25)$$

The Eq. (25) means that, if the step size satisfies $0 < \gamma \leq 1$, $L(\gamma) \leq \frac{1}{\gamma}$ will be satisfied then pFISTA used to solve the proposed model in Eq. (8) will convergence with a speed described in **Theorem 1.**

## IV. EXPERIMENTAL RESULTS

In this section, we describe our experiments with three *in vivo* liver DCE datasets to demonstrate that the proposed method produces better clinical visual image quality and faster reconstruction speed compared to two state-of-the-art reference DCE reconstruction methods including K-T ARTS-GROWL [12] and iGRASP [13]. For different methods, several parameters must be set firstly. The step size in the proposed method is set to be 1 for fast convergence speed. Reconstruction parameters for all methods (TABLE I) were empirically selected to obtain good image quality without introducing significant temporal blurring. Because fully-sampled images are not available in DCE-MRI for validation, four radiologists (with 25, 23, 21, 26 years of clinical experience in abdominal imaging, respectively), who are blind to the reconstruction methods visually assessed the overall image quality of *in vivo* DCE-MRI, considering the sharpness of vessel, residual artifacts and noise.

TABLE I
RECONSTRUCTION PARAMETERS

| Dataset ID | iGRASP | K-T ARTS-GROWL | Proposed |
|---|---|---|---|
| 1 | $\lambda_i = 0.04 M_0$ | $\lambda_k = 0.01 P_0$ | $\lambda_p = 0.04 M_0$, $w = 0.2$ |
| 2 | $\lambda_i = 0.05 M_0$ | $\lambda_k = 0.02 P_0$ | $\lambda_p = 0.06 M_0$, $w = 0.09$ |
| 3 | $\lambda_i = 0.03 M_0$ | $\lambda_k = 0.015 P_0$ | $\lambda_p = 0.025 M_0$, $w = 0.3$ |

Note: $P_0$ is the maximal magnitude of the PI results, $M_0$ is the maximal magnitude of the NUFFT results, also used to initialize iGRASP and the proposed method reconstruction.

Three liver DCE datasets were downloaded from http://cai2r.net/resources/software. The first and the second datasets were acquired on a 3 Tesla MRI Scanners (Siemens AG, Erlangen, Germany) with a 12-channel coil array. A three-dimensional radial stack-of-stars fast low-angle shot (FLASH) pulse sequence with golden-angle reordering scheme was employed for free-breathing data acquisitions. Relevant imaging parameters of the first dataset were: FOV = 380 * 380 mm$^2$, TR/TE = 3.9/1.7 ms, Partitions = 30, Slice Thickness = 3 mm, Spokes in Each Partition = 600, Sampling in Each Readout = 384, Acquisition Time = 77 s. For the second dataset: FOV = 370 * 370 mm$^2$, TR/TE = 3.83/1.71 ms, Partitions = 40, Slice Thickness = 3 mm, Spokes in Each Partition = 600, Sampling in Each Readout = 768, Acquisition Time = 90s. The last dataset was acquired with a fat-saturated stack-of-stars golden-angle radial imaging sequence relevant parameters were: FOV = 350 * 350 mm$^2$, TR/TE = 3.6/1.6 ms, Partitions = 48, Slice Thickness = 5 mm, Spokes in Each Partition = 1100, Sampling in Each Readout = 512, Acquisition Time = 190 s.

All reconstruction tasks were implemented in MATLAB 2018a (Mathworks Inc, Natick, MA) on a personal computer with 2.80 GHz dual-core CPU and 8 GB RAM.

*A. Main results*

Representative partitions from three reconstructed DCE-MR images are shown in Figs. 5-7 with zoomed views of different regions of interest (ROIs). For display purpose, only three contrast phases that are most relevant to clinical diagnosis, including one pre-contrast phase, one arterial phase, and one venous phase, are presented. The visual image quality scores listed in TABLE II, suggesting that the proposed method achieved significantly better visual image quality (p values: iGRASP vs proposed: 3.37e-12, K-T ARTS-GROWL vs proposed: 2.31e-49, calculated with the Wilcoxon signed-rank sum test where p < 0.05 was considered to be statistically significant difference). All radiologists all made their consensus that the reconstruction image series of the proposed method present better overall quality, higher sharpness, improved delineation of hepatic vessels and reduced level of noise and artifacts.

The aorta, highlighted with a blue circle in Figs. 5-7, was selected to evaluate temporal fidelity [12, 13] because it is a region showing the highest variation of signal in DCE liver images [40]. The aorta signal intensity time courses of three different methods reconstruction results and corresponding reference were averaged and shown in Fig. 8. It can be seen that all the methods show similar enhancement patterns. For K-T ARTS-GROWL, parallel imaging alone was used as reference [12], and for iGRASP, NUFFT results were used as reference [13]. To determine the accuracy of the curves, linear correlation was evaluated. The linear correlations between K-T ARTS-GROWL and the reference are both larger than 0.99 for all three datasets. iGRASP and the proposed method results are shown in Fig. 9, both of which are larger than 0.99 close to 1. Thus, the proposed method does not introduce significant temporal blurring and it offered a comparable temporal resolution with the other two state-of-the-art methods.

The reconstruction time listed in TABLE III implies the fast reconstruction among all methods.

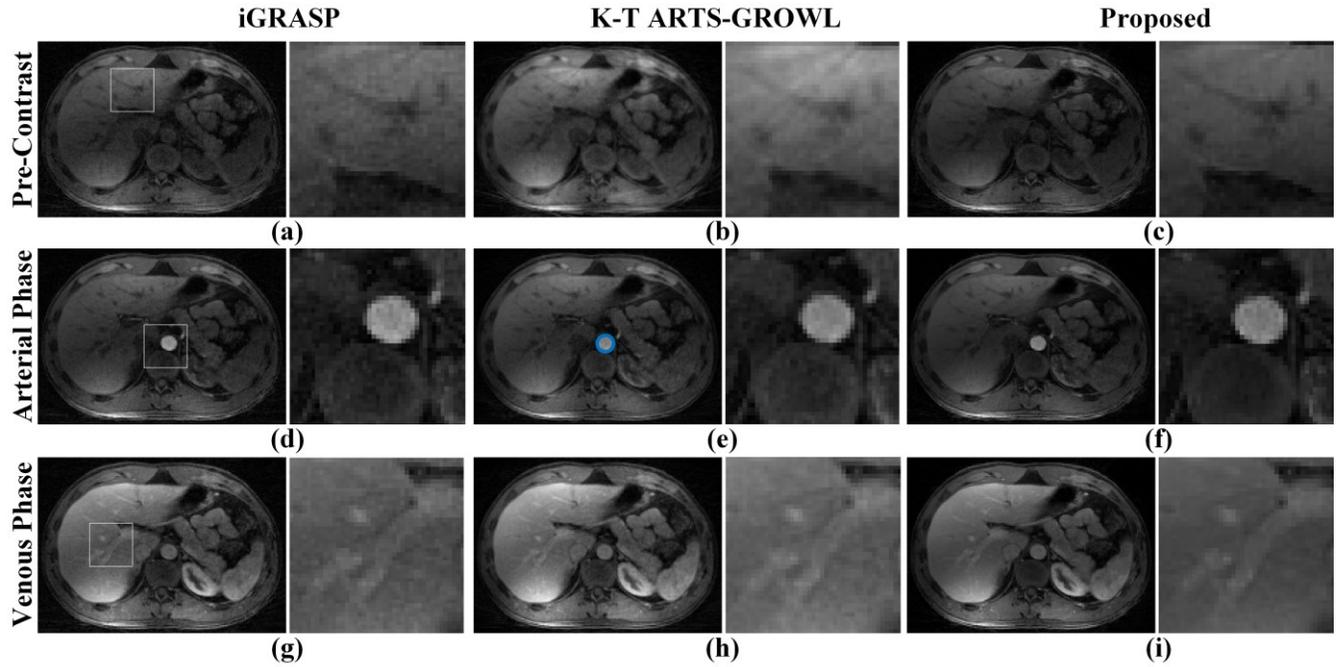

Fig. 5. Reconstructed three frames of the representative partition in dataset 1, iGRASP in the first column, K-T ARTS-GROWL in the second, the proposed method in the last by grouping 21 consecutive spokes to form each temporal frame with a temporal resolution as 21 spokes per frame, 28 frames in total. The reconstructed matrix of each frame is 192*192. The ROIs were called out by white squares for comparisons.

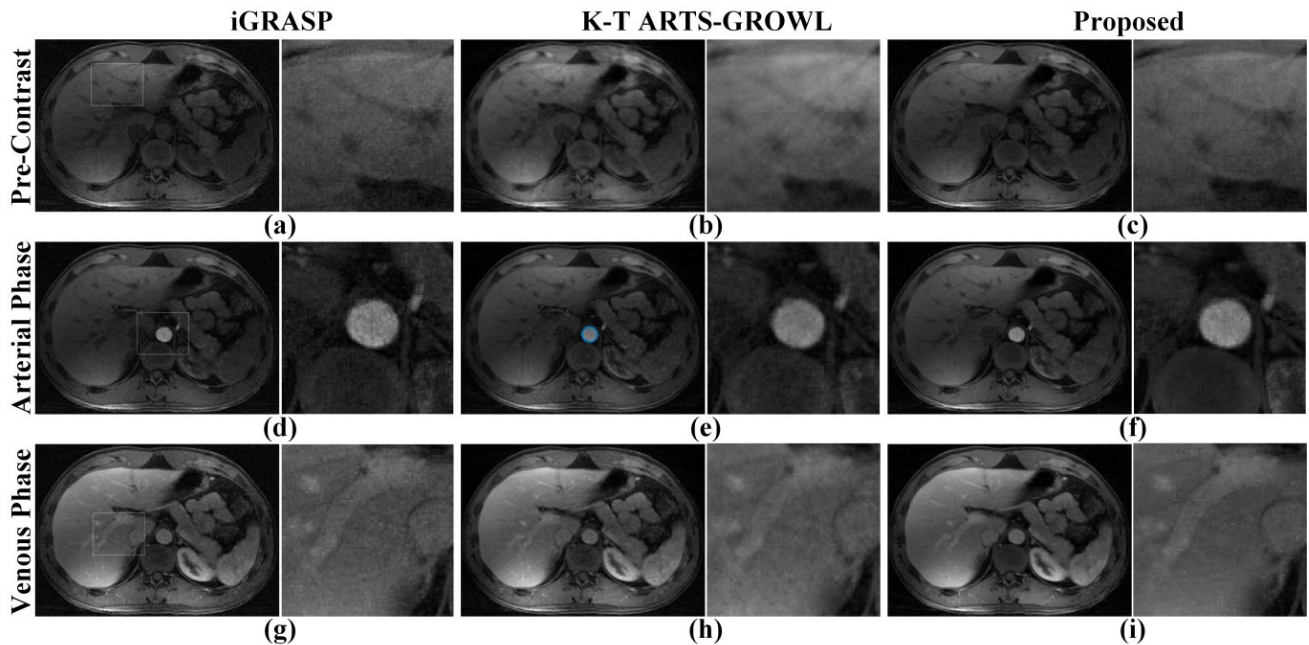

Fig. 6. Reconstructed three frames of the representative partition in dataset 2, iGRASP in the first column, K-T ARTS-GROWL in the second, the proposed method in the last by grouping 21 consecutive spokes to form each temporal frame with a temporal resolution as 21 spokes per frame, 28 frames in total. The reconstructed matrix of each frame is 384*384. The ROIs were called out by white squares for comparisons.

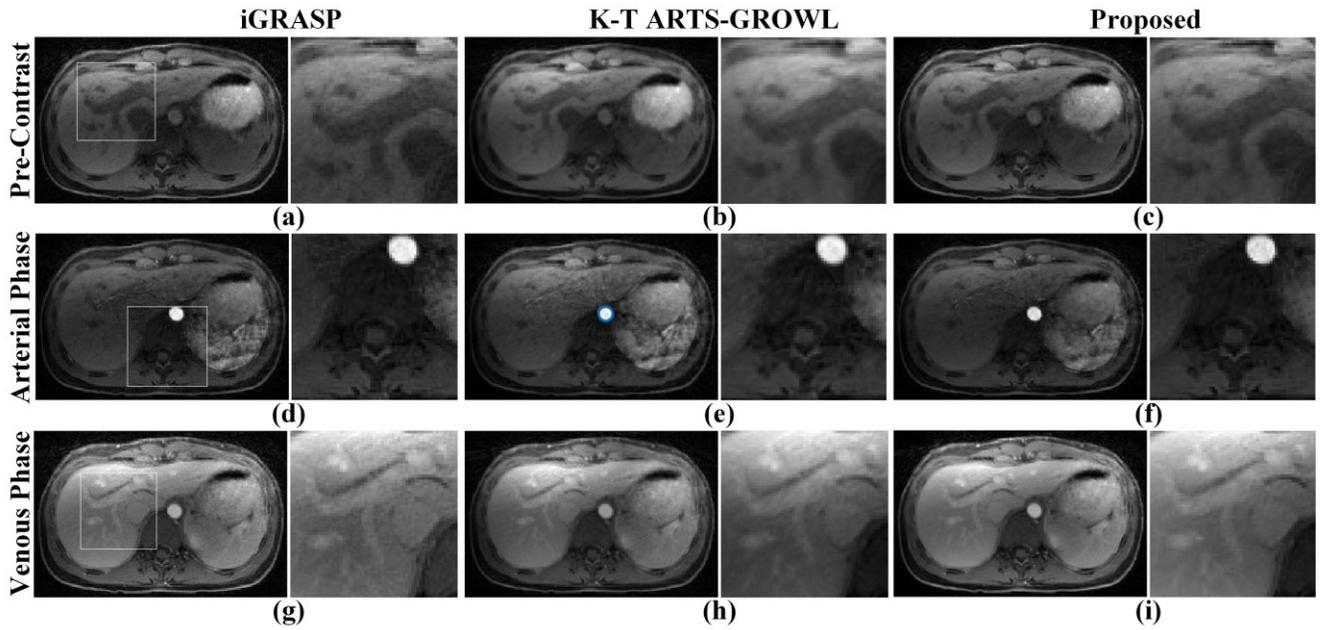

Fig. 7. Reconstructed three frames of the representative partition in dataset 3, iGRASP in the first column, K-T ARTS-GROWL in the second, the proposed method in the last by grouping 34 consecutive spokes to form each temporal frame with a temporal resolution as 34 spokes per frame, 32 frames in total. The reconstructed matrix of each frame is 256*256. The ROIs were called out by white squares for comparisons.

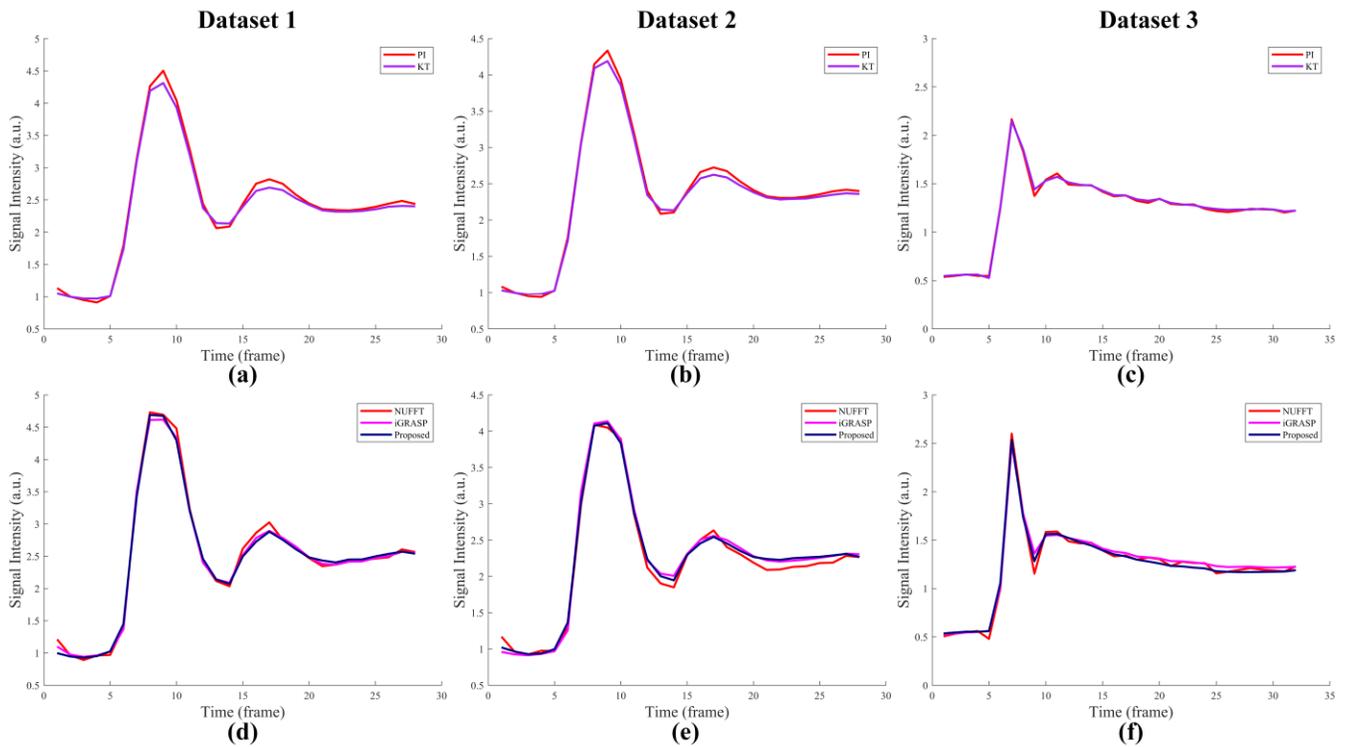

Fig. 8. Aorta enhancement curves of different methods. (a)-(c) KT ARTS-GROWL (KT) results, (d)-(f) iGRASP and the proposed method results, where the proposed method used NUFFT results as reference like iGRASP.

TABLE II
IMAGE QUALITY ASSESSMENT

| Dataset ID | iGRASP | KT ARTS-GROWL | Proposed |
|---|---|---|---|
| 1 | 3.19 ± 0.88 | 2.91 ± 1.26 | **3.51 ± 1.15** |
| 2 | 3.31 ± 0.99 | 3.03 ± 1.09 | **3.75 ± 1.13** |
| 3 | 3.27 ± 1.22 | 2.61 ± 1.14 | **3.34 ± 1.19** |

The visual image quality scores are display as means ± standard deviations. Scoring rule: 5 = excellent, 4 = good, 3 = adequate, 2 = poor, 1 = non-diagnostic.

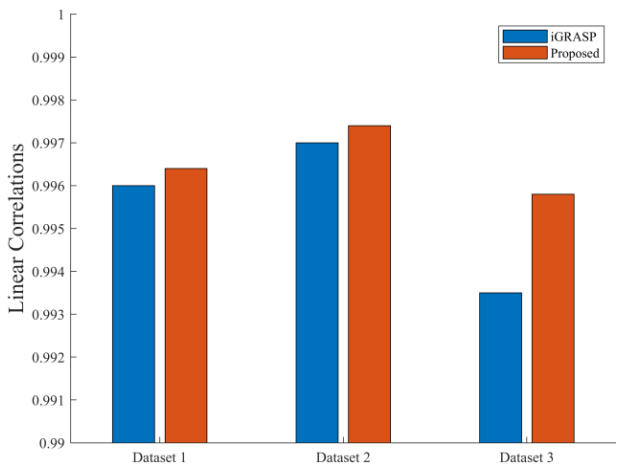

Fig. 9. The linear correlations of the signal intensity time courses between iGRASP and NUFFT, proposed method and NUFFT in three different datasets, where the correlation larger than 0.99 was considered to be no significant temporal blurring.

TABLE III
RECONSTRUCTION TIME (S)

| Dataset ID | iGRASP | K-T ARTS-GROWL | Proposed |
|---|---|---|---|
| 1 | 1118 | 1764 | **183** |
| 2 | 1881 | 6316 | **738** |
| 3 | 1430 | 5867 | **568** |

Note: the gridding time is counted in K-T ARTS-GROWL.

## V. CONCLUSION

In this work, we proposed a tight frame-based compressed sensing reconstruction model with flexible weighted sparse constraint along both time and spatial dimension for parallel DCE-MRI. We modified the pFISTA, a fast and efficient algorithm, to solve the reconstruction model and theoretically analyze the convergence. Experimental results demonstrate that the proposed method outperforms the two state-of-the-art DCE-MRI reconstruction methods on whatever the clinical image quality assessment and the reconstruction speed, while maintaining comparable temporal resolution. The proposed method may also be extended to other dynamic MRI or imaging modality applications.

ACKNOWLEDGMENTS

The authors would like to thank the Center for Advanced Imaging Innovation and Research (CAI$^2$R) at New York University School of Medicine for sharing the DCE datasets used in this paper. The authors are grateful to Zhifeng Chen for sharing the K-T ARTS-GROWL codes. The authors appreciate the help of Xianfeng Chen, and Khan Afsar for polishing the language of this paper.

**Appendix**

In this appendix, we will explain how to use CSDWT and SIDWT [1-5] to sparsify the DCE image series in time and spatial dimensions separately. The DCE image series $\mathbf{d}$ rearranged to a column vector can be expressed as:

$$\mathbf{d} = [\mathbf{x}_1; \mathbf{x}_2; \ldots; \mathbf{x}_J] \in \mathbb{C}^{NJ}, \quad (A1)$$

where $\mathbf{x}_j \in \mathbb{C}^N (j=1,2,\ldots,J)$ is the $j^{th}$ frame vector of the image series.

1. CSDWT along the time dimension

The pipeline of CSDWT and inverse CSDWT used to DCE image series $\mathbf{d}$ along time dimension are shown in Fig. 1.

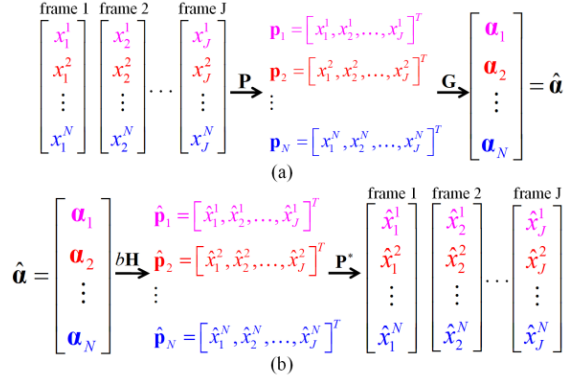

Fig. 1. (a) CSDWT, $x_j^n$ is the $n^{th}$ pixel of frame j in $\mathbf{d}$; (b) Inverse CSDWT, $\hat{x}_j^n$ is the $n^{th}$ pixel of frame j in $\hat{\mathbf{d}}$.

The mainly three steps when we adopt CSDWT to sparsify $\mathbf{d}$ along time dimension contain:

(1) Extracting pixels of $\mathbf{d}$ corresponding to the time dimension:

$$\mathbf{p} = [\mathbf{P}_1; \mathbf{P}_2; \ldots; \mathbf{P}_N]\mathbf{d} = \mathbf{Pd} = [\mathbf{p}_1; \mathbf{p}_2; \ldots; \mathbf{p}_N] \in \mathbb{C}^{NJ}, \quad (A2)$$

where $\mathbf{P} = [\mathbf{P}_1; \mathbf{P}_2; \ldots; \mathbf{P}_N] \in \mathbb{R}^{NJ \times NJ}$, $\mathbf{P}_n \in \mathbb{R}^{J \times NJ} (n=1,2,\ldots,N)$ denotes extracting the $n^{th}$ pixel of each frame in $\mathbf{d}$, $\mathbf{p}_n = \mathbf{P}_n\mathbf{d} \in \mathbb{C}^J (n=1,2,\ldots,N)$ is the $n^{th}$ pixels of each frame in $\mathbf{d}$.

(2) Cyclic shifting the pixels in $\mathbf{p}_n$ for $M$ times ($M=6$ in this work) then we can get:

$$\mathbf{c}_n = [\mathbf{C}_0; \mathbf{C}_1; \ldots; \mathbf{C}_M]\mathbf{p}_n = \mathbf{C}\mathbf{p}_n \in \mathbb{C}^{(M+1)J}, \quad (A3)$$

where $\mathbf{C} = [\mathbf{C}_0; \mathbf{C}_1; \ldots; \mathbf{C}_M] \in \mathbb{R}^{(M+1)J \times J}$, $\mathbf{C}_m \in \mathbb{R}^{J \times J} (m=0,1,\ldots,M)$, denotes cyclic shifting the pixels in $\mathbf{p}_n$ to the right for $m$ times (one pixel a time in this work).

(3) Discrete wavelet transform (DWT)

$$\boldsymbol{\alpha}_n = \begin{bmatrix} \mathbf{D} & & 0 \\ & \ddots & \\ 0 & & \mathbf{D} \end{bmatrix} \mathbf{c}_n = \tilde{\mathbf{D}}\mathbf{c}_n \in \mathbb{C}^{(M+1)J}, \quad (A4)$$

where $\mathbf{D} \in \mathbb{C}^{J \times J}$ denotes DWT, Daubechies wavelets with 5 decomposition levels are utilized in this work, $\tilde{\mathbf{D}} = \begin{bmatrix} \mathbf{D} & & 0 \\ & \ddots & \\ 0 & & \mathbf{D} \end{bmatrix} \in \mathbb{C}^{(M+1)J \times (M+1)J}$.

Then we can get the sparse coefficients of the DCE image series after CSDWT along time dimension:

$$\hat{\boldsymbol{\alpha}} = [\boldsymbol{\alpha}_1; \boldsymbol{\alpha}_2; \ldots; \boldsymbol{\alpha}_N] = [\tilde{\mathbf{D}}\mathbf{c}_1; \tilde{\mathbf{D}}\mathbf{c}_2; \ldots; \tilde{\mathbf{D}}\mathbf{c}_N] = \mathbf{GPd} \in \mathbb{C}^{N(M+1)J}, \quad (A5)$$

where: $\mathbf{G} = \begin{bmatrix} \tilde{\mathbf{D}}\mathbf{C} & & 0 \\ & \ddots & \\ 0 & & \tilde{\mathbf{D}}\mathbf{C} \end{bmatrix} \in \mathbb{C}^{N(M+1)J \times NJ}$, then the CSDWT can be expressed as: $\mathbf{R}_T = \mathbf{GP}$.

Inverse CSDWT:

(1) Applying inverse DWT to $\boldsymbol{\alpha}_n$, then we can get:

$$\tilde{\boldsymbol{\alpha}}_n = \tilde{\mathbf{D}}^*\boldsymbol{\alpha}_n = \tilde{\mathbf{D}}^*\tilde{\mathbf{D}}\mathbf{c}_n \in \mathbb{C}^{(M+1)J}, \quad (A6)$$

where $\tilde{\mathbf{D}}^* = \begin{bmatrix} \mathbf{D}^* & & 0 \\ & \ddots & \\ 0 & & \mathbf{D}^* \end{bmatrix} \in \mathbb{C}^{(M+1)J \times (M+1)J}$, $\mathbf{D}^*$ is the adjoint of $\mathbf{D}$, and it satisfies $\mathbf{D}^*\mathbf{D} = \mathbf{I}$.

(2) Inverse shifting of the pixels in $\tilde{\boldsymbol{\alpha}}_n$:

$$\hat{\mathbf{c}}_n = \hat{\mathbf{C}}\tilde{\boldsymbol{\alpha}}_n = \hat{\mathbf{C}}\tilde{\mathbf{D}}^*\tilde{\mathbf{D}}\mathbf{c}_n \in \mathbb{C}^{(M+1)J}, \quad (A7)$$

where $\hat{\mathbf{C}} = \begin{bmatrix} \hat{\mathbf{C}}_0 & & & 0 \\ & \hat{\mathbf{C}}_1 & & \\ & & \ddots & \\ 0 & & & \hat{\mathbf{C}}_M \end{bmatrix} \in \mathbb{R}^{(M+1)J \times (M+1)J}$, $\hat{\mathbf{C}}_m \in \mathbb{R}^{J \times J} (m=0,1,\ldots,M)$, denotes cyclic shifting the pixels in $\tilde{\boldsymbol{\alpha}}_n$ to the left for $m$ times (one pixel a time), and it satisfies $\hat{\mathbf{C}}_m\mathbf{C}_m = \mathbf{I}$.

(3) Sum and average of the pixels

Denote: $\hat{\mathbf{I}} = [\mathbf{I} \ \mathbf{I} \ \cdots \ \mathbf{I}] \in \mathbb{R}^{J \times J(M+1)}$, $\mathbf{I} \in \mathbb{R}^{J \times J}$ is an identity matrix, let: $b = (M+1)^{-1}$, therefore: $\hat{\mathbf{p}}_n = b\hat{\mathbf{I}}\hat{\mathbf{c}}_n = b\hat{\mathbf{I}}\hat{\mathbf{C}}\tilde{\mathbf{D}}^*\tilde{\mathbf{D}}\mathbf{c}_n \in \mathbb{C}^J$, then we can get:

$$\hat{\mathbf{p}} = [\hat{\mathbf{p}}_1; \hat{\mathbf{p}}_2; \ldots; \hat{\mathbf{p}}_N] = b[\hat{\mathbf{I}}\hat{\mathbf{C}}\tilde{\mathbf{D}}^*\tilde{\mathbf{D}}\mathbf{c}_1; \hat{\mathbf{I}}\hat{\mathbf{C}}\tilde{\mathbf{D}}^*\tilde{\mathbf{D}}\mathbf{c}_2; \ldots; \hat{\mathbf{I}}\hat{\mathbf{C}}\tilde{\mathbf{D}}^*\tilde{\mathbf{D}}\mathbf{c}_n]$$
$$= b \begin{bmatrix} \hat{\mathbf{I}}\hat{\mathbf{C}}\tilde{\mathbf{D}}^* & & 0 \\ & \ddots & \\ 0 & & \hat{\mathbf{I}}\hat{\mathbf{C}}\tilde{\mathbf{D}}^* \end{bmatrix} \hat{\boldsymbol{\alpha}} \in \mathbb{C}^{NJ}. \quad (A8)$$

Denote: $\mathbf{H} = \begin{bmatrix} \hat{\mathbf{I}}\hat{\mathbf{C}}\tilde{\mathbf{D}}^* & & 0 \\ & \ddots & \\ 0 & & \hat{\mathbf{I}}\hat{\mathbf{C}}\tilde{\mathbf{D}}^* \end{bmatrix} \in \mathbb{C}^{NJ \times NJ(M+1)}$, $\hat{\mathbf{d}} = \mathbf{P}^*\hat{\mathbf{p}} = b\mathbf{P}^*\mathbf{H}\hat{\boldsymbol{\alpha}} \in \mathbb{C}^{NJ}$, so the inverse CSDWT can be expressed as: $\mathbf{R}_T^* = b\mathbf{P}^*\mathbf{H}$.

2. SIDWT in spatial dimension

$$\boldsymbol{\alpha}_S = \begin{bmatrix} \boldsymbol{\Psi} & & 0 \\ & \ddots & \\ 0 & & \boldsymbol{\Psi} \end{bmatrix} \mathbf{d} = [\boldsymbol{\Psi}\mathbf{x}_1; \boldsymbol{\Psi}\mathbf{x}_2; \ldots; \boldsymbol{\Psi}\mathbf{x}_J], \quad (A9)$$

where $\boldsymbol{\alpha}_S$ is the sparse coefficients after applying SIDWT in spatial dimension, $\boldsymbol{\Psi}$ denotes SIDWT, Daubechies wavelets with 4 decomposition levels are utilized in this work. So, the tight frame adopted to sparsify $\mathbf{d}$ in spatial dimension can be expressed as: $\mathbf{R}_S = \begin{bmatrix} \boldsymbol{\Psi} & & 0 \\ & \ddots & \\ 0 & & \boldsymbol{\Psi} \end{bmatrix}$, inverse SIDWT in spatial dimension is $\mathbf{R}_S^* = \begin{bmatrix} \boldsymbol{\Psi}^* & & 0 \\ & \ddots & \\ 0 & & \boldsymbol{\Psi}^* \end{bmatrix}$, where $\boldsymbol{\Psi}^*$ is the adjoint of $\boldsymbol{\Psi}$.